\documentclass{article}
\usepackage{latexsym}
\newtheorem{corollary}{Corollary}
\newtheorem{definition}{Definition}
\newtheorem{lemma}{Lemma}
\newtheorem{theorem}{Theorem}
\begin{document}

\title{PPT from spectra}

\author{Roland Hildebrand \thanks{%
LMC, Universit\'e Joseph Fourier, Tour IRMA, 51 rue des Math\'ematiques, 38400 St.\ Martin d'H\`eres, France
({\tt roland.hildebrand@imag.fr}). This paper presents research results of the Action Concert\'ee Incitative
"Masses de donn\'ees" of CNRS, France. The scientific responsibility rests with its author.}}

\maketitle

\begin{abstract}
In this contribution we solve the following problem. Let $H_{nm}$ be a Hilbert space of dimension $nm$, and
let $A$ be a positive semidefinite self-adjoint linear operator on $H_{nm}$. Under which conditions on the spectrum
has $A$ a positive partial transpose (is PPT) with respect to any partition $H_n \otimes H_m$ of the space $H_{nm}$ as a
tensor product of an $n$-dimensional and an $m$-dimensional Hilbert space? We show that the necessary
and sufficient conditions can be expressed as a set of linear matrix inequalities (LMIs) on the
eigenvalues of $A$.
\end{abstract}

\section{Introduction}

This paper is motivated by a problem posed by Emanuel Knill and listed as Problem 15 in the list of open problems in Quantum Information Theory
on the website of the Institute of Mathematical Physics at the TU Braunschweig \cite{openprobs}. Knill posed the following question.

\smallskip

{\it Suppose $A$ is a self-adjoint PSD operator on a Hilbert space $H_{nm}$ of dimension $N = nm$. Which are the conditions on the {\sl spectrum} of $A$
guaranteeing that $A$ is {\sl separable} with respect to any decomposition of $H_{nm}$ as a tensor product $H_n \otimes H_m$ of two Hilbert
spaces of dimensions $n,m$, respectively? }

\smallskip

The problem was solved for the case $n=m=2$ (a 2-qubit system) by Verstraete et al.\ \cite{Verstraete01}, who showed that a necessary and sufficient condition
is represented by a quadratic inequality on the eigenvalues of $A$.

It is well-known that a necessary condition for separability is the so-called {\sl positive partial transpose} (PPT) condition \cite{Woronowicz},\cite{Peres96}. Therefore
any condition on the spectrum of $A$ that is necessary for the PPT property to hold with respect to any decomposition $H_{nm} = H_n \otimes H_m$ will also be necessary
for separability with respect to any decomposition.

\smallskip

In this contribution we explicite necessary and sufficient conditions on the spectrum under which $A$ exhibits the PPT property (is PPT) with respect to arbitrary decompositions
of fixed dimension. These conditions are expressed as linear matrix inequalities (LMIs) on the eigenvalues of $A$. Since the PPT property is equivalent to separability for the cases
$m=2$, $n=2,3$ \cite{Woronowicz}, we solve the original problem of E.\ Knill for these two cases and thus furnish an exact solution for a second special case.
For arbitrary dimensions the obtained LMIs are necessary conditions for separability. The number of LMIs depends only on the minimum $\min(n,m)$ of the factor dimensions.

Another result that emerged from our study is that the property of being PPT for arbitrary decompositions of $H_N$ into a tensor product of spaces of fixed dimensions $n,m$
gets stronger if $\min(n,m)$ gets bigger. In other words, let $N = nm = n'm'$ with $\min(n,m) \geq \min(n',m')$. If a PSD operator $A$ is PPT with respect to arbitrary decompositions
$H_N = H_n \otimes H_m$, then it is also PPT with respect to arbitrary decompositions $H_N = H_{n'} \otimes H_{m'}$.

The remainder of the paper is structured as follows. In the next section we derive the necessary and sufficient conditions on the spectrum of $A$ in the form of LMIs.
In Section 3 we show the above-mentioned dependence of the stringence of the condition on the dimensions of the decomposition. In the fourth section we illustrate
our results for the cases $\min(n,m)=2,3$ and list the corresponding LMI conditions explicitely. In the last section we summarize our results.

In the sequel $H_n$ denotes a Hilbert space of dimension $n$. Let $A$ be a hermitian matrix of size $nm \times nm$. We consider $A$ as consisting of $m \times m$ blocks of size
$n \times n$ each. The {\sl partial transpose} of $A$, denoted by $A^{\Gamma}$, will be defined as an $nm \times nm$-matrix consisting of the same blocks, but with the blocks $(k,l)$, $(l,k)$ 
interchanged for $1 \leq k < l \leq m$.

\section{Main theorem}

In this section we derive a set of LMIs on the spectrum of a self-adjoint PSD operator $A$ on $H_{nm}$ which represents a necessary and sufficient condition for $A$ to have the PPT property 
with respect to any decomposition of $H_{nm}$ as a tensor product $H_n \otimes H_m$.

\smallskip

Let $x \in {\bf R}^n$ be an arbitrary real vector of dimension $n$. Let us define a set $E(x)$ by
\[ E(x) = \{x_k^2,\ (k = 1,\dots,n);\ +x_kx_l,-x_kx_l,\ (1 \leq k < l \leq n)\}.
\]
This set contains $n^2$ real numbers.

Our main result is based on the following lemma.

{\lemma \label{eigpart} Let $B = bb^*$ be a hermitian PSD rank 1 matrix of size $nm$. Here $b$ is a vector in $H_{nm}$.
Then there exists a real vector $x \in {\bf R}^p$ with non-negative, ordered entries $x_1 \geq x_2 \geq \cdots \geq x_p \geq 0$,
where $p = \min(n,m)$, such that the spectrum of the partial transpose $B^{\Gamma}$ is given by the set $E(x)$, the remaining $p|n-m|$ eigenvalues being zero.
The elements $x_k$, $k = 1,\dots,p$ are the singular values of the $n \times m$ matrix ${\bf b}$ that is obtained from $b$ by
arranging the $m$ $n$-dimensional subvectors $b_1,\dots,b_m$ of $b$ columnwise. }

{\it Remark:} Obviously the assertion of the lemma holds also if $B = 0$.

{\it Proof.} Let $B = bb^*$ satisfy the assumptions of the lemma, let $b_1,\dots,b_m$ be the $n$-dimensional subvectors of $b$
and let ${\bf b}$ be the $n \times m$ matrix composed of these subvectors.

Let now ${\bf b} = U_n D V_m$ be the singular value decomposition of the matrix ${\bf b}$, where $U_n,V_m$
are unitary operators of appropriate sizes and $D$ a diagonal matrix of size $n \times m$ containing the singular values
$x_1,\dots,x_p$ of ${\bf b}$ in decreasing order.

Then $W = U_n^* \otimes \bar V_m$, $W_{\Gamma} = U_n^* \otimes V_m$ are unitary operators on $H_{nm}$. It is easily seen that
the relation $(WBW^*)^{\Gamma} = W_{\Gamma}B^{\Gamma}W_{\Gamma}^*$ holds. In particular, the spectrum of $B^{\Gamma}$ equals the spectrum of $(WBW^*)^{\Gamma}$.

Let us determine the structure of $(WBW^*)^{\Gamma}$. Let $v_1,\dots,v_m$ be the rows of $V_m$. It is not hard
to see that
\[ w = Wb = (U_n^* \otimes \bar V_m)\left( \begin{array}{c} b_1 \\ \vdots \\ b_m \end{array} \right) = \left(
\begin{array}{c} U_n^*{\bf b}v_1^* \\ \vdots \\ U_n^*{\bf b}v_m^* \end{array} \right) = vec(D),
\]
where the operator $vec$ stacks the columns of the matrix it is applied to into a big column vector. Hence
the vector $w = Wb$ has $x_k$ as $(k+(k-1)n)$-th element, $k = 1,\dots,p$, and all other elements are zero.
Partition the matrix $WBW^* = ww^*$ into $m \times m$ blocks of size $n \times n$. The block at
position $(k,l)$ ($k,l \leq p$) has the product $x_kx_l$ at position $(k,l)$, all other elements being zero.

Let us consider the partial transpose $(WBW^*)^{\Gamma} = (ww^*)^{\Gamma}$ of this matrix. Its block at
position $(k,l)$ has the product $x_kx_l$ at position $(l,k)$ ($k,l \leq p$), all other elements being zero. It is not hard
to see that this matrix can be block-diagonalised by a permutation of the rows and columns, with diagonal
blocks
\[ x_k^2, (k = 1,\dots,p);\ \left( \begin{array}{cc} 0 & x_kx_l \\ x_kx_l & 0 \end{array} \right),\ (1 \leq k <
l \leq p),
\]
all other blocks being zero. Therefore the spectrum of the matrix $(WBW^*)^{\Gamma}$ is given by $x_k^2$, $k =
1,\dots,p$; $+x_kx_l,-x_kx_l$, $1 \leq k < l \leq p$, the rest of the eigenvalues being zero. The first $p^2$ numbers
form exactly the set $E(x)$, where $x$ is the vector composed of the singular values $x_1,\dots,x_p$ of ${\bf b}$.
This completes the proof of the lemma. $\Box$

{\lemma \label{xtob} Let $n,m$ be positive integers and let $p = \min(n,m)$. Let further $x \in {\bf R}^p$ be
an arbitrary vector. Then there exists a vector $b \in H_{nm}$ such that the spectrum of $(bb^*)^{\Gamma}$ is given by the set $E(x)$, the remaining $p|n-m|$ eigenvalues being zero. }

{\it Proof.} Define $b$ elementwise as follows. Let $x_k$ be the $(k+(k-1)n)$-th element of $b$, $k = 1,\dots,p$,
and let all other elements be zero. Let the matrix ${\bf b}$, as in the previous lemma, be composed
columnwise of the $m$ $n$-dimensional subvectors $b_1,\dots,b_m$ of $b$. Then ${\bf b}$ is diagonal with $x_k$
as diagonal elements. Hence its singular values are the absolute values
$|x_k|$. Applying Lemma \ref{eigpart}, we get that the eigenvalues of
$(bb^*)^{\Gamma}$ are given by the products $|x_k|^2$, $k = 1,\dots,p$; $+|x_k| |x_l|,-|x_k| |x_l|$, $1 \leq k
< l \leq p$, the rest of the eigenvalues being zero. But this is exactly the spectrum we claimed
$(bb^*)^{\Gamma}$ to possess. $\Box$

{\lemma \label{commutemin} Let $A,B$ be hermitian matrices of size $n \times n$. Let $a_1,\dots,a_n$
and $b_1,\dots,b_n$ be their eigenvalues in decreasing order. Then
\[ \min_{U\ {\mbox {\rm unitary}}} \langle UAU^*, B \rangle = \sum_{k=1}^n a_{n+1-k}b_k.
\] }

{\it Proof.} The scalar product $\langle UAU^*, B \rangle$ is an analytic function $f(U)$ of the unitary matrix $U$,
which ranges over a compact set. Let $U_0$ realize the minimum $f^*$ of this function. Then the first order
extremality condition states that for any skew-hermitian matrix $S$ we have $\langle SU_0AU_0^*, B \rangle +
\langle U_0AU_0^*S^*, B \rangle = \langle S, [U_0AU_0^*,B] \rangle = 0$. In other words, the commutator
$[U_0AU_0^*,B]$ is hermitian. Since the commutator of two hermitian matrices is always skew-hermitian, the
matrices $U_0AU_0^*$ and $B$ must commute. But then we can diagonalize them simultaneously by conjugation
with some unitary matrix $V$. Moreover, we have $\langle U_0AU_0^*, B \rangle = \langle VU_0AU_0^*V^*, VBV^*
\rangle$.

Therefore the minimum $f^*$ is given by $\min_{\sigma \in S_n} \sum_{k=1}^n a_{\sigma(k)}b_k$, where $\sigma$
ranges over all permutations of the indices $1,\dots,n$. This minimum is attained at the
inversion $\sigma^*$ defined by $\sigma^*(k) = n+1-k$. Indeed, let $\sigma \in S_n$ and $k < l$ be such that $\sigma(k) < \sigma(l)$. Then $a_{\sigma(k)} \leq a_{\sigma(l)}$ and
$b_k \leq b_l$. It follows that $(a_{\sigma(k)}-a_{\sigma(l)})(b_k-b_l) \geq 0$ and hence
$a_{\sigma(k)}b_k + a_{\sigma(l)}b_l \geq a_{\sigma(l)}b_k + a_{\sigma(k)}b_l$. Therefore an interchange of
$\sigma(k)$ and $\sigma(l)$ can only decrease the value of the sum $\sum_{k=1}^n a_{\sigma(k)}b_k$.
Performing this interchange consecutively for all pairs $(k,l)$ for which $k < l$ and $\sigma(k) < \sigma(l)$, we finally arrive at the inversion $\sigma^*$,
regardless of the permutation $\sigma$ we started with. This completes the proof. $\Box$

\smallskip

Let $A$ be a self-adjoint PSD operator on $H_{nm}$. Let
$\lambda_1,\dots,\lambda_{nm}$ be the eigenvalues of $A$ in decreasing order and let $p = \min(n,m)$. Define
also $p_+ = p(p+1)/2$, $p_- = p(p-1)/2$ and let $S_+ = \{(k,l) \,|\, 1 \leq k \leq l \leq p\}$, $S_- =
\{(k,l) \,|\, 1 \leq k < l \leq p\}$ be sets of index pairs with cardinalities $p_+,p_-$, accordingly. Note
that we have $E(x) = \{x_kx_l \,|\, (k,l) \in S_+\} \cup \{ -x_kx_l \,|\, (k,l) \in S_-\}$ for $x \in {\bf
R}^p$. In the sequel we will consider {\sl orderings} of the sets $S_+,S_-$. We define an ordering of a
finite set $S$ as a bijective map $\sigma$ from $S$ onto the set $\{1,2,\dots,\#S\}$, where $\#S$ is the
cardinality of $S$.

Let $x \in {\bf R}^p$ be a vector with non-negative entries.

{\definition We say that an ordering $\sigma_+: S_+ \to \{1,\dots,p_+\}$ of $S_+$ is {\sl compatible} with $x$
if for any two index pairs $(k_1,l_1),(k_2,l_2) \in S_+$ such that $\sigma_+(k_1,l_1) < \sigma_+(k_2,l_2)$ 
we have $x_{k_1}x_{l_1} \geq x_{k_2}x_{l_2}$. 

We say that a pair of orderings $(\sigma_+: S_+ \to \{1,\dots,p_+\}, \sigma_-: S_- \to
\{1,\dots,p_-\})$ of the sets $S_+,S_-$ is {\sl compatible} with $x$ if $\sigma_+$ is compatible with $x$ and 
for any two index pairs $(k_1,l_1),(k_2,l_2) \in S_-$ such that $\sigma_+(k_1,l_1) <
\sigma_+(k_2,l_2)$ (recall that $S_- \subset S_+$) we have $\sigma_-(k_1,l_1) <
\sigma_-(k_2,l_2)$. }

Thus for a compatible ordering 1 is the image of the index pair $(k,l) \in S_{\pm}$ for which the product $x_kx_l$ is largest, 2 the
image of the pair for which $x_kx_l$ is second-largest and so on. Note that for any $x$ there exists at least one pair of
orderings which is compatible with $x$.

{\corollary \label{charac} The operator $A$ is PPT for all decompositions of $H_{nm}$ as a tensor product
space $H_n \otimes H_m$ if and only if for any vector $x \in {\bf R}^p$ with non-negative and ordered entries
$x_1 \geq x_2 \geq \cdots \geq x_p \geq 0$ there exists a pair $(\sigma_+,\sigma_-)$ of orderings which is
compatible with $x$ such that
\begin{equation} \label{rawsum}
\sum_{(k,l) \in S_+} \lambda_{nm+1-\sigma_+(k,l)} x_kx_l - \sum_{(k,l) \in S_-} \lambda_{\sigma_-(k,l)}
x_kx_l \geq 0.
\end{equation} }

{\it Proof.} By definition, $A$ is PPT for all decompositions of $H_{nm}$ if for all unitary matrices $U$ the
partial transpose $(UAU^*)^{\Gamma}$ is PSD. This is the case if for all vectors $b \in H_{nm}$ we have
\[ b^* (UAU^*)^{\Gamma} b = \langle (UAU^*)^{\Gamma}, bb^* \rangle = \langle UAU^*, (bb^*)^{\Gamma} \rangle \geq
0.
\]
Let us fix $b$ for the moment. By Lemma \ref{eigpart} there exist non-negative numbers $x_1 \geq \cdots \geq
x_p$ such that the spectrum of $(bb^*)^{\Gamma}$ consists of the set $E(x)$, the rest of the eigenvalues being
zero. Note that, apart from the zeros, there are $p_+$ non-negative eigenvalues $x_kx_l$, $(k,l) \in S_+$,
and $p_-$ non-positive eigenvalues $-x_kx_l$, $(k,l) \in S_-$. If $(\sigma_+,\sigma_-)$ is any pair of
orderings compatible with $x$, then we have by Lemma \ref{commutemin}
\[ \min_U b^* (UAU^*)^{\Gamma} b = \min_U \langle UAU^*, (bb^*)^{\Gamma} \rangle = \sum_{(k,l) \in S_+}
\lambda_{nm+1-\sigma_+(k,l)} x_kx_l - \sum_{(k,l) \in S_-} \lambda_{\sigma_-(k,l)} x_kx_l.
\]
If the expression on the right-hand side is non-negative for any vector $x$ with non-negative and ordered
entries, then the expression on the left-hand side is non-negative for all vectors $b$. Hence in this case
$A$ is PPT for all decompositions of $H_{nm}$.

Let now $x \in {\bf R}^p$ be a vector with non-negative entries, and $(\sigma_+,\sigma_-)$ a pair of
orderings which is compatible with $x$. Suppose that
\[ \sum_{(k,l) \in S_+}
\lambda_{nm+1-\sigma_+(k,l)} x_kx_l - \sum_{(k,l) \in S_-} \lambda_{\sigma_-(k,l)} x_kx_l < 0.
\]
By Lemmas \ref{xtob} and \ref{commutemin} there exists a vector $b \in H_{nm}$ such that
\[ \min_U b^* (UAU^*)^{\Gamma} b = \sum_{(k,l) \in S_+}
\lambda_{nm+1-\sigma_+(k,l)} x_kx_l - \sum_{(k,l) \in S_-} \lambda_{\sigma_-(k,l)} x_kx_l < 0.
\]
Let $U_0$ be the unitary matrix that realizes this minimum. Then we get that $U_0AU_0^*$ is not PPT. Thus if
$A$ is PPT for all decompositions of $H_{nm}$, then inequality (\ref{rawsum}) holds for any vector $x$ with
non-negative entries and any pair of orderings $(\sigma_+,\sigma_-)$ which is compatible with $x$. But such
pairs do exist for any vector $x$. This completes the proof. $\Box$

\smallskip

Let us transform expression (\ref{rawsum}). Let $(\sigma_+,\sigma_-)$ be a pair of orderings. Define a $p
\times p$-matrix $\Lambda(\sigma_+,\sigma_-)$ elementwise as follows.
\[ \Lambda_{kl}(\sigma_+,\sigma_-) = \left\{ \begin{array}{lll} \lambda_{nm+1-\sigma_+(k,l)}, & \quad & k \leq l, \\
-\lambda_{\sigma_-(l,k)}, && k > l. \end{array} \right.
\]
Then we have
\begin{equation} \label{eqcharac}
\sum_{(k,l) \in S_+}
\lambda_{nm+1-\sigma_+(k,l)} x_kx_l - \sum_{(k,l) \in S_-} \lambda_{\sigma_-(k,l)} x_kx_l = x^T
\Lambda(\sigma_+,\sigma_-) x.
\end{equation}
The matrix $\Lambda$ has thus signed eigenvalues of $A$ as elements, namely the $p_+$ smallest eigenvalues in the
upper triangular part, including the diagonal, and the $p_-$ largest eigenvalues with a minus sign in the lower triangular
part. The particular arrangement depends on the orderings $\sigma_+$, $\sigma_-$. Define the finite set of pairs
\[ \Sigma_{\pm} = \left\{ (\sigma_+,\sigma_-) \,|\, \exists\ x_1 > x_2 > \cdots > x_p > 0:\ (\sigma_+,\sigma_-)\
\mbox{compatible with}\ x = (x_1,\dots,x_p)^T \right\}.
\]

{\lemma \label{existsigma} Let $x \in {\bf R}^p$ be a vector with non-negative ordered entries $x_1 \geq
\cdots \geq x_p \geq 0$. Then there exists a pair $(\sigma_+,\sigma_-) \in \Sigma_{\pm}$ which is
compatible with $x$. }

{\it Proof.} Let $x$ satisfy the assumptions of the lemma.
Then there exists a sequence of vectors $x^k \in {\bf R}^p$ such that $\lim_{k \to \infty} x^k = x$ and the components
of $x^k$ satisfy the inequalities $x_1^k > x_2^k > \cdots > x_p^k > 0$ for all $k$. By the definition of $\Sigma_{\pm}$,
for any $k$ there exists a pair of orderings $(\sigma_+^k,\sigma_-^k) \in \Sigma_{\pm}$ that is compatible with $x^k$.
Let $(\sigma_+,\sigma_-)$ be an accumulation point of the sequence $\{(\sigma_+^k,\sigma_-^k)\}$. Obviously
$(\sigma_+,\sigma_-)$ is in $\Sigma_{\pm}$ and compatible with $x$. $\Box$

\smallskip

Now we are ready to prove our main theorem.

{\theorem \label{main} The operator $A$ is PPT for all decompositions of $H_{nm}$ as a tensor product space $H_n \otimes
H_m$ if and only if for all $(\sigma_+,\sigma_-) \in \Sigma_{\pm}$ we have
\[ \Lambda(\sigma_+,\sigma_-) + \Lambda(\sigma_+,\sigma_-)^T \succeq 0.
\] }

{\it Proof.} Suppose that $\Lambda(\sigma_+,\sigma_-) + \Lambda(\sigma_+,\sigma_-)^T \succeq 0$ for all
$(\sigma_+,\sigma_-) \in \Sigma_{\pm}$. Then $x^T \Lambda(\sigma_+,\sigma_-) x \geq 0$ for all $x \in {\bf
R}^p$ and for all $(\sigma_+,\sigma_-) \in \Sigma_{\pm}$. From Corollary \ref{charac}, relation
(\ref{eqcharac}) and Lemma \ref{existsigma} it follows that $A$ is PPT for all decompositions of $H_{nm}$.

Let now $(\sigma_+,\sigma_-) \in \Sigma_{\pm}$ and $x \in {\bf R}^p$ be such that $x^T
\Lambda(\sigma_+,\sigma_-) x < 0$. We shall show that there exists a unitary matrix $U$ such that $UAU^*$ is
not PPT. By Lemma \ref{xtob} there exists a vector $b \in H_{nm}$ such that the matrix $(bb^*)^{\Gamma}$
has the spectrum $E(x)$, the rest of the eigenvalues being zero. Denote the eigenvalues of $(bb^*)^{\Gamma}$
by $\mu_1,\dots,\mu_{nm}$.

The expression on the left-hand side of (\ref{eqcharac}) is of the form
\[ \sum_{k=1}^{nm} \lambda_{\sigma(k)} \mu_k,
\]
where $\sigma$ is a permutation of the index set $\{1,\dots,nm\}$. There exist unitary matrices $U,V$ such
that $UAU^* = diag(\lambda_{\sigma(1)},\lambda_{\sigma(2)},\dots,\lambda_{\sigma(nm)})$, $V (bb^*)^{\Gamma} V^*
= diag(\mu_1,\dots,\mu_{nm})$. Hence
\[ x^T \Lambda(\sigma_+,\sigma_-) x = \langle UAU^*, V(bb^*)^{\Gamma}V^* \rangle = \langle
(V^*UAU^*V)^{\Gamma}, bb^* \rangle < 0,
\]
and $V^*UAU^*V$ is not PPT. This completes the proof. $\Box$

\smallskip

Thus we have expressed the sought necessary and sufficient conditions on the spectrum of $A$ as a set of
LMIs.

\section{Decompositions of different dimensions}

In this section we prove that if $N = n_1m_1 = n_2m_2$ with $\min(n_1,m_1) \geq \min(n_2,m_2)$, than any self-adjoint
PSD operator $A$ on $H_N$ which is PPT for any decomposition $H_N = H_{n_1} \otimes H_{m_1}$, is also PPT for any
decomposition $H_N = H_{n_2} \otimes H_{m_2}$.

In the sequel we explicite the dependence on the dimension $p$ of the sets $S_+,S_-,\Sigma_{\pm}$ defined in the previous section, 
i.e.\ we write $S_+(p),S_-(p),\Sigma_{\pm}(p)$. Our result is based on the following lemma.

{\lemma Let $p > q$ and suppose that the elements of the vector $y \in {\bf R}^q$ satisfy the inequalities
$y_1 > y_2 > \cdots > y_q > 0$. Let $(\sigma_+,\sigma_-)$ be a pair of orderings of the sets $S_+(q),S_-(q)$ that is compatible with $y$.
Then there exists a vector $x \in {\bf R}^p$ with elements satisfying the inequalities $x_1 > x_2 > \cdots
> x_p > 0$ and a pair of orderings $(\rho_+,\rho_-)$ of the sets $S_+(p),S_-(p)$ which is compatible with $x$ such that $\rho_+$
equals $\sigma_+$ on the domain $S_+(q)$ of definition of $\sigma_+$ and $\rho_-$ equals $\sigma_-$ on the domain $S_-(q)$ of
definition of $\sigma_-$. }

{\it Proof.} Let $y$ and $(\sigma_+,\sigma_-)$ satisfy the assumptions of the lemma. Define the vector $x$ as
follows. For $k \leq q$ let $x_k = y_k$. For $k > q$ choose $x_k$ such that $\frac{y_q^2}{y_1}
> x_{q+1} > x_{q+2} > \cdots > x_p > 0$.

Since $\frac{y_q}{y_1} < 1$, we have $x_q = y_q > \frac{y_q^2}{y_1} > x_{q+1}$. Therefore the sequence
$\{x_k\}$ is strictly decreasing. Further, let $(k,l) \in S_+(q)$ and $(k',l') \in S_+(p) \setminus S_+(q)$. Then we have $l' > q$. It follows that $x_{k'}x_{l'} \leq
x_1x_{q+1} < x_1\frac{y_q^2}{y_1} = x_q^2 \leq x_kx_l$. Therefore $\rho_+(k',l') > \rho_+(k,l)$ for any
ordering $\rho_+$ that is compatible with $x$. As a consequence, any such ordering maps $S_+(q)$ onto the set $\{1,\dots,\#S_+(q)\}$. 
Let $\rho_+$ be an ordering of $S_+(p)$ which is compatible with $x$. Define another ordering $\rho'_+$ of $S_+(p)$ by
\[ \rho'_+(k,l) = \left\{ \begin{array}{rcl} \sigma_+(k,l), & \quad & (k,l) \in S_+(q), \\
\rho_+(k,l), & & (k,l) \in S_+(p) \setminus S_+(q). \end{array} \right.
\]
Since $\sigma_+$ is compatible with $y$, and $x_k = y_k$ for $k \leq q$, we have also that $\rho'_+$ is compatible with $x$.
Now it rests to choose $\rho'_-$ as the unique ordering of $S_-(p)$ that makes the pair $(\rho'_+,\rho'_-)$ compatible with $x$. $\Box$

\smallskip

Now we are ready to prove the above-mentioned result.

{\theorem \label{dims} Let $N = n_1m_1 = n_2m_2$ with $p_1 = \min(n_1,m_1) \geq p_2 = \min(n_2,m_2)$. Let $A$ be a self-adjoint
PSD operator on $H_N$ which is PPT with respect to any decomposition $H_N = H_{n_1} \otimes H_{m_1}$. Then $A$ is also PPT with respect to any
decomposition $H_N = H_{n_2} \otimes H_{m_2}$. }

{\it Proof.} Assume the conditions of the theorem. Let $(\sigma_+,\sigma_-) \in \Sigma_{\pm}(p_2)$. By the preceding lemma there exists a pair of orderings $(\rho_+,\rho_-) \in \Sigma_{\pm}(p_1)$
such that $\rho_+$ equals $\sigma_+$ on $S_+(p_2)$ and $\rho_-$ equals $\sigma_-$ on $S_-(p_2)$. It follows that the upper left $p_2 \times p_2$-subblock of the matrix
$\Lambda(\rho_+,\rho_-)$ equals the matrix $\Lambda(\sigma_+,\sigma_-)$. But $\Lambda(\rho_+,\rho_-) + \Lambda(\rho_+,\rho_-)^T \succeq 0$
by Theorem \ref{main} and the assumption on $A$. Hence we have also $\Lambda(\sigma_+,\sigma_-) + \Lambda(\sigma_+,\sigma_-)^T \succeq 0$. 
Application of Theorem \ref{main} completes the proof. $\Box$

\section{Examples}

In this section we illustrate the obtained results on the examples of $2 \times n$ and $3 \times n$ bipartite
spaces.

\smallskip

Let $m = 2$, $n \geq 2$. Then we have $p = 2$, $p_+ = 3$, $p_- = 1$. Let us find the set $\Sigma_{\pm}(2)$. If
$x \in {\bf R}^2$ is a vector with elements $x_1 > x_2 > 0$, then we have the relations $x_1^2 > x_1x_2 > x_2^2$ on the
products $x_kx_l$, $(k,l) \in S_+(2)$. Hence the only element of $\Sigma_{\pm}(2)$ is the pair
$(\sigma_+,\sigma_-)$ defined by
\[ \sigma_+: \left\{ \begin{array}{rcl} (1,1) & \mapsto & 1, \\
(1,2) & \mapsto & 2, \\ (2,2) & \mapsto & 3, \end{array} \right. \quad \sigma_-: (1,2) \mapsto 1.
\]
The corresponding matrix $\Lambda$ is given by
\[ \Lambda(\sigma_+,\sigma_-) = \left( \begin{array}{ccc} \lambda_{2n} & \lambda_{2n-1} \\ -\lambda_1 &
\lambda_{2n-2} \end{array} \right).
\]
Hence we obtain the following necessary and sufficient LMI condition:
\[ \left( \begin{array}{ccc} 2\lambda_{2n} & \lambda_{2n-1}-\lambda_1 \\ \lambda_{2n-1}-\lambda_1 &
2\lambda_{2n-2} \end{array} \right) \succeq 0.
\]
Since $\lambda_k \geq 0$ by the positivity of $A$, and $\lambda_1 \geq \lambda_{2n-1}$, this matrix inequality reduces to
\[ 4\lambda_{2n}\lambda_{2n-2} - (\lambda_{2n-1}-\lambda_1)^2 \geq 0 \ \Leftrightarrow\ \lambda_1 \leq
\lambda_{2n-1} + 2\sqrt{\lambda_{2n}\lambda_{2n-2}}.
\]

We obtain the following corollary.

{\corollary \label{m2} Let $A$ be a self-adjoint PSD operator on the Hilbert space $H_{2n}$. Let $\lambda_1,\dots,\lambda_{2n}$ be the
eigenvalues of $A$ in decreasing order. Then $A$ is PPT with respect to any decomposition of $H_{2n}$ as a tensor product
$H_2 \otimes H_n$ if and only if the inequality $\lambda_1 \leq \lambda_{2n-1} + 2\sqrt{\lambda_{2n}\lambda_{2n-2}}$ holds. $\Box$ }

\smallskip

For $n = 2$ and $n=3$ this inequality takes the forms
\begin{equation} \label{2x2}
\lambda_1 \leq \lambda_3 + 2\sqrt{\lambda_2\lambda_4},
\end{equation}
\begin{equation} \label{2x3}
\lambda_1 \leq \lambda_5 + 2\sqrt{\lambda_4\lambda_6}.
\end{equation}
It is well-known \cite{Woronowicz} that for the cases $m=2, n=2,3$ the PPT condition is equivalent to separability. Therefore we obtain also
the following corollary.

{\corollary \label{sep} Let $A$ be a self-adjoint PSD operator on the Hilbert space $H_4$ ($H_6$). Let $\lambda_1,\lambda_2,\lambda_3,\lambda_4$ ($\lambda_1,\dots,\lambda_6$) be the
eigenvalues of $A$ in decreasing order. Then $A$ is separable with respect to any decomposition of $H_4$ ($H_6$) as a tensor product
$H_2 \otimes H_2$ ($H_2 \otimes H_3$) if and only if inequality (\ref{2x2}) ((\ref{2x3})) holds. $\Box$ }

Inequality (\ref{2x2}) is already well-known to be a necessary and sufficient condition for separability for any partition of $H_4$ as $H_2
\otimes H_2$ and was presented as partial solution of Knill's {\sl Problem 15} in quantum information theory \cite{Verstraete01}. We present here
inequality (\ref{2x3}) as partial solution for the case $m=2$, $n=3$.

\smallskip

Let now $m = 3$, $n \geq 3$. Then $p = 3$, $p_+ = 6$, $p_- = 3$. We shall now determine the set $\Sigma_{\pm}(3)$. If $x_1 > x_2 > x_3 > 0$, then we have
\[ x_1^2 > x_1x_2 > \max(x_2^2,x_1x_3) \geq \min(x_2^2,x_1x_3) > x_2x_3 > x_3^2.
\]
However, we can have both $x_2^2 \geq x_1x_3$ and $x_1x_3 \geq x_2^2$. Hence $\Sigma_{\pm}(3)$ consists of two
elements, and the corresponding matrices $\Lambda$ are given by
\[ \Lambda_1 = \left( \begin{array}{ccc} \lambda_{3n} & \lambda_{3n-1} & \lambda_{3n-3} \\ -\lambda_1 &
\lambda_{3n-2} & \lambda_{3n-4} \\ -\lambda_2 & -\lambda_3 & \lambda_{3n-5} \end{array} \right), \quad
\Lambda_2 = \left( \begin{array}{ccc} \lambda_{3n} & \lambda_{3n-1} & \lambda_{3n-2} \\ -\lambda_1 &
\lambda_{3n-3} & \lambda_{3n-4} \\ -\lambda_2 & -\lambda_3 & \lambda_{3n-5} \end{array} \right).
\]
Thus we obtain the two LMIs
\begin{equation} \label{3xn}
\left( \begin{array}{ccc} 2\lambda_{3n} & \lambda_{3n-1}-\lambda_1 & \lambda_{3n-3}-\lambda_2 \\ \lambda_{3n-1}-\lambda_1 &
2\lambda_{3n-2} & \lambda_{3n-4}-\lambda_3 \\ \lambda_{3n-3}-\lambda_2 & \lambda_{3n-4}-\lambda_3 &
2\lambda_{3n-5} \end{array} \right) \succeq 0, \quad \left( \begin{array}{ccc} 2\lambda_{3n} &
\lambda_{3n-1}-\lambda_1 &
\lambda_{3n-2}-\lambda_2 \\ \lambda_{3n-1}-\lambda_1 & 2\lambda_{3n-3} & \lambda_{3n-4}-\lambda_3 \\
\lambda_{3n-2}-\lambda_2 & \lambda_{3n-4}-\lambda_3 & 2\lambda_{3n-5}
\end{array} \right) \succeq 0.
\end{equation}

{\corollary Let $A$ be a self-adjoint PSD operator on the Hilbert space $H_{3n}$. Let $\lambda_1,\dots,\lambda_{3n}$ be the
eigenvalues of $A$ in decreasing order. Then $A$ is PPT with respect to any decomposition of $H_{3n}$ as a tensor product
$H_3 \otimes H_n$ if and only if linear matrix inequalities (\ref{3xn}) hold. $\Box$ }

\section{Summary}

In this contribution we presented necessary and sufficient conditions on the spectrum of a self-adjoint positive semidefinite operator $A$ on a Hilbert space
$H_{nm}$ of dimension $nm$ under which $A$ has a positive partial transpose for any decomposition of $H_{nm}$ as a tensor product space
$H_n \otimes H_m$, where $H_n,H_m$ are Hilbert spaces of dimensions $n,m$. These conditions have the form of linear matrix inequalities on the
eigenvalues of $A$ (Theorem \ref{main}). We showed that if these conditions hold for a pair of dimensions $(n,m)$, then they hold also for a pair $(n',m')$ (where $nm = n'm'$)
whenever $\min(n',m') \leq \min(n,m)$ (Theorem \ref{dims}). Hence the condition is most stringent if $n \approx m$.

For the case $\min(n,m) = 2$ we reduced the LMI condition to a single inequality (Corollary \ref{m2}). For the cases $m=2$, $n=2,3$ our conditions
are necessary and sufficient also for {\it separability} of $A$ with respect to an arbitrary decomposition of the underlying space $H_4$ or $H_6$ (Corollary \ref{sep}).
While the result for $n=m=2$ was known before, the result for $m=2$, $n=3$ is new and provides a solution of E.\ Knills open problem number 15 \cite{openprobs} 
for another special case.

\end{document}